\pdfoutput=1

\documentclass[aps, prd
, onecolumn
, 11pt
, tightenlines
, nofootinbib 
, superscriptaddress
]{revtex4-2}
\usepackage{graphicx}
\usepackage{xcolor}
\usepackage[caption=false]{subfig}
\usepackage{mathrsfs,mathtools}
\usepackage{
amssymb}
\usepackage{physics}
\usepackage{bm}
\usepackage{braket}
\usepackage{listings}
\usepackage{cases}
\usepackage{comment}
\usepackage{soul}
\usepackage{cancel}
\usepackage{cases}
\usepackage[utf8]{inputenc}
\usepackage{url}
\usepackage{longtable}
\usepackage{xspace}
\usepackage{aas_macros}
\usepackage[colorlinks=true
,urlcolor=darkblue
,anchorcolor=darkblue
,citecolor=darkblue
,filecolor=darkblue
,linkcolor=darkblue
,menucolor=darkblue
,linktocpage=true
,pdfproducer=medialab
,pdfa=true
]{hyperref}

\newcommand{\ee}{\mathrm{e}}

\newcommand{\Mpl}{M_\mathrm{Pl}}
\newcommand{\ns}{n_{\mathrm{s}}}

\newcommand{\fNL}{f_\mathrm{NL}}

\newcommand{\USR}{\mathrm{USR}}
\newcommand{\SR}{\mathrm{SR}}
\newcommand{\kL}{k_\mathrm{L}}

\newcommand{\eff}{\mathrm{eff}}

\newcommand{\uc}{\mathrm{c}}

\newcommand{\uf}{\mathrm{f}}

\newcommand{\ui}{\mathrm{i}}

\newcommand{\bfk}{\mathbf{k}}
\newcommand{\uL}{\mathrm{L}}

\newcommand{\calN}{\mathcal{N}}
\newcommand{\calO}{\mathcal{O}}

\newcommand{\uS}{\mathrm{S}}
\newcommand{\ut}{\mathrm{t}}
\newcommand{\bfx}{\mathbf{x}}
\newcommand{\bfy}{\mathbf{y}}

\newcommand{\bae}[1]{\begin{align} #1 \end{align}}
\newcommand{\bce}[1]{\begin{cases} #1 \end{cases}}
\newcommand{\dps}{\displaystyle}
\newcommand{\bfe}[4]{
\begin{figure} 
	\centering
	\includegraphics[#1]{#2}
	\caption{#3}
	\label{#4}
\end{figure}}

\newcommand{\relmiddle}[1]{\mathrel{}\middle#1\mathrel{}}

\definecolor{monza}{HTML}{CF000F}
\definecolor{darkblue}{HTML}{00008b}
\definecolor{darkmagenta}{HTML}{8b008b}

\begin{document}
\title{Revisiting non-Gaussianity in non-attractor inflation models in the light of the cosmological soft theorem}
\date{\today}

\author{Teruaki Suyama}
\affiliation{Department of Physics, Tokyo Institute of Technology, 2-12-1 Ookayama, Meguro-ku, Tokyo 152-8551, Japan}
\author{Yuichiro Tada}
\email{tada.yuichiro@e.mbox.nagoya-u.ac.jp}
\affiliation{Department of Physics, Nagoya University, Nagoya 464-8602, Japan}
\author{Masahide Yamaguchi}
\affiliation{Department of Physics, Tokyo Institute of Technology, 2-12-1 Ookayama, Meguro-ku, Tokyo 152-8551, Japan}

\begin{abstract}
We revisit the squeezed-limit non-Gaussianity in the single-field non-attractor inflation models from the viewpoint of the cosmological soft theorem. In the single-field attractor models, inflaton's trajectories with different initial conditions effectively converge into a single trajectory in the phase space, and hence there is only one \emph{clock} degree of freedom (DoF) in the scalar part. 
Its long-wavelength perturbations can be absorbed into the local coordinate renormalization and lead to the so-called \emph{consistency relation} between $n$- and $(n+1)$-point functions. On the other hand, if the inflaton dynamics deviates from the attractor behavior, its long-wavelength perturbations cannot necessarily be absorbed and the consistency relation is expected not to hold any longer. In this work, we derive a formula for the squeezed bispectrum including the explicit correction to the consistency relation, as a proof of its violation in the non-attractor cases. 
First one must recall that non-attractor inflation needs to be followed by attractor inflation in a realistic case. 
Then, even if a specific non-attractor phase is effectively governed by a single DoF of phase space (represented by the exact ultra-slow-roll limit) and followed by a single-DoF attractor phase, its transition phase necessarily involves two DoF in dynamics and hence its long-wavelength perturbations cannot be absorbed into the local coordinate renormalization. Thus, it can affect local physics, even taking account of the so-called \emph{local observer effect}, as shown by the fact that the bispectrum in the squeezed limit can go beyond the consistency relation. 
More concretely, the observed squeezed bispectrum does not vanish in general for long-wavelength perturbations exiting the horizon during a non-attractor phase.
\end{abstract}

\maketitle

\section{Introduction}

The primordial curvature perturbations give us rich information on the dynamics of an inflation as well as serve as a seed of large scale structure formation. Their statistical features can be probed even beyond the linear order recently. Then, the soft limit of the correlation functions of the primordial curvature perturbations becomes a powerful tool to probe the inflation dynamics through clarifying the relation between their correlation functions. One of such examples is so-called Maldacena's consistency relation~\cite{Maldacena:2002vr}, which connects the squeezed limit of the bispectrum ($(n+1)$-point functions) to the power spectra ($n$-point functions). This relation holds true~\cite{Creminelli:2004yq,Cheung:2007sv,Hui:2018cag}, as long as (i) there is only a single scalar field showing the attractor behavior (single clock inflation) and (ii) the vacuum is the Bunch-Davies one~\cite{Agullo:2010ws,Berezhiani:2014kga}. In this paper, we focus on the former assumption~(i). It should be noticed that, even if there is only a single scalar field, Maldacena's consistency relation can be violated. Such an example is ultra-slow-roll (USR) inflation (a non-attractor inflation)~\cite{Tsamis:2003px,Kinney:2005vj}. In USR inflation, its dynamics shows non-attractor-like behavior, depending solely on its momentum (velocity) thanks to the shift symmetry (though trajectories do not converge because of the dependence of an initial field value), and the comoving curvature perturbations can evolve even after the horizon exit, which might lead to the violation of the consistency relation~\cite{Namjoo:2012aa,Martin:2012pe}.
This feature is in sharp contrast to that of standard slow-roll (attractor) inflation, where the momentum of a scalar field is determined by the field value, that is, trajectories with different initial conditions effectively converge into a single trajectory in the phase space and the curvature perturbations become constants after the horizon exit. In this case, its long-wavelength curvature perturbations can be renormalized into the redefinition of the local (background) coordinate, which leads to the consistency relation.

Recently, there appear some discrepancies among the literature~\cite{Cai:2017bxr,Passaglia:2018ixg,Bravo:2017gct,Bravo:2017wyw,Bravo:2020hde} on the soft limit of primordial curvature perturbations for an inflation model including a non-attractor phase. USR inflation cannot last long enough to solve the horizon and flatness problems without fine-tuning of an initial condition and cannot fit the observed density perturbations, so that it needs to be followed by standard slow-roll (SR) inflation. Then, the key question is whether Maldacena's consistency relation for a long-wavelength mode exiting the horizon during an USR phase is finally recovered after an USR phase ends and the comoving curvature perturbations stop growing.\footnote{The non-Gaussianity in USR models is important also in the context of the primordial black hole (see, e.g., Refs.~\cite{Byrnes:2012yx,Passaglia:2018ixg,Ragavendra:2020sop}).}
Some literature~\cite{Cai:2017bxr,Passaglia:2018ixg} demonstrates that the final bispectrum in the squeezed limit depends on the feature of the transition from an USR phase to a SR phase and can violate the consistency relation. On the other hand, another~\cite{Bravo:2020hde} claims that it finally reduces to Maldacena's consistency relation. In particular, in Refs.~\cite{Bravo:2017gct,Bravo:2017wyw,Bravo:2020hde}, they extended Maldacena's consistency relation to that including the time evolution of the curvature perturbations so as to discuss the case of USR inflation. According to their formula, they claim that, once an USR phase ends and the comoving curvature perturbations stop growing, it would finally recover Maldacena's consistency relation.\footnote{Ref.~\cite{Finelli:2017fml} derives the same formula as Refs.~\cite{Bravo:2017gct,Bravo:2017wyw,Bravo:2020hde}. However they explicitly restrict their discussion to theories with the exact shift symmetry, having no conflict with our results.}

In this paper, we revisit this issue and point out that it is not enough to include the time dependence of the comoving curvature perturbations to discuss the above issue. In fact, one needs to further extend the soft-limit bispectrum formula given in Refs.~\cite{Bravo:2017gct,Bravo:2017wyw,Bravo:2020hde} by including the momentum dependence of an inflaton in order to deal with an inflation model including a non-attractor phase. This is because, as we mentioned before, though the dynamics of an inflaton during both of USR and SR phases are governed by a single degree of freedom and its long-wavelength perturbations can be absorbed into the local coordinate renormalization, two phase space degrees of freedom necessarily play an important role in the dynamics during the \emph{transition} period, which leave imprints on a later SR phase. Thus, we need to take momentum dependence into account for the evolution of the comoving curvature perturbations even after the renormalization of field-value dependence, which can finally lead to the violation of Maldacena's consistency relation for an inflation model including a non-attractor phase, reproducing the results in the literature successfully.  

We also mention the local observer effects~\cite{Tanaka:2011aj,Pajer:2013ana,Tada:2016pmk}.\footnote{The local observer effects on another soft limit called collapsed limit~\cite{Suyama:2007bg,Suyama:2010uj} are discussed in Ref.~\cite{Suyama:2020akr}.} As first noticed in Ref.~\cite{Tanaka:2011aj}, a local observer cannot be the comoving observer due to the modulation of the long wavelength mode. In case of standard slow-roll and single-field inflation, this local observer effect exactly cancels the squeezed limit of the bispectrum and hence the observed bispectrum in that limit vanishes except the secondary effects. Then, 
the question arises what happens for USR inflation. Some literature~\cite{Bravo:2017gct,Bravo:2017wyw,Bravo:2020hde} claims that the observed squeezed bispectrum universally vanishes irrespective of attractor or non-attractor inflation models. In this paper, we also try to address this question and will show that the observed bispectrum in the squeezed limit for a long-wavelength mode exiting the horizon during an USR phase does not vanish in general.  

The paper is organized as follows. In the next section, we derive the long-wavelength modulation on the short mode by including not only time dependence but also momentum dependence in order to deal with a non-attractor inflation model (with a transition to a SR phase) as well as a pure attractor one. Then, we give the generalized single-field soft theorem. In Sec.~\ref{Sec:Bis}, we demonstrate a generic method to evaluate the squeezed bispectrum of the curvature perturbations by taking account of the transition from an USR phase to a SR phase adequately. In Sec.~\ref{Sec:Exa}, concrete examples are discussed to show how powerful our formula is and to reproduce the results obtained in the literature. 
The final section is devoted to conclusions and discussion

\section{Consistency relation or not in non-attractor cases}
\label{Sec:Con}

Maldacena's first computation of the cubic action reveals that the following consistency relation holds between the power spectrum $P_\zeta$ and the squeezed bispectrum $B_\zeta$ of the curvature perturbation $\zeta$ in attractor single-field inflation~\cite{Maldacena:2002vr}:
\bae{\label{eq: Maldacena CR}
    B_\zeta(q,k,k^\prime)\underset{q\ll k,k^\prime}{\to}(1-\ns)P_\zeta(k)P_\zeta (q),
}
where $\ns$ is the spectral index defined by
\bae{
    \ns-4=\dv{\log P_\zeta}{\log k}.
}
It is now well understood as one example of the so-called cosmological soft theorem (see, e.g., Refs.~\cite{Huang:2006eha,2012JCAP...08..001S}), which connects the squeezed $(n+1)$-point function with the $n$-point function.

The ordinary cosmological soft theorem is a consequence of the fact that the long-wavelength perturbation in the unique scalar degree of freedom (DoF) can be locally absorbed into the spacetime coordinate renormalization in the single-field and attractor cases.
It is easily understood in the comoving gauge, where the inflaton field $\phi$ has no perturbation,
\bae{
    \phi(t,\bfx)=\phi_0(t),
}
from the background value $\phi_0(t)$ and all perturbations are expressed in the metric form. 
The separate universe assumption requires that
the spacetime geometry is approximated by the flat FLRW one 
in the long-wavelength limit as
\bae{\label{eq: local FLRW}
    \dd{s}^2=-\dd{t}^2+\bar{a}^2(t)\dd{\bfx}^2+\calO\pqty{\frac{\kL^2}{a^2H^2}}.
}
Here the local scale factor $\bar{a}=a\ee^{\zeta_\uL}$ is modulated from the background $a$ by the long-wavelength curvature perturbation $\zeta_\uL$.
$H=\dot{a}/a$ is the Hubble parameter and $k_\uL$ is the corresponding wavenumber to the superhorizon perturbation $\zeta_\uL$.
The general conservation of the curvature perturbation on a superhorizon scale ensures that $\zeta_\uL$ is constant in the attractor case~\cite{Lyth:2004gb}, and any constant renormalization of the scale factor does not affect the local physics.
The only difference is the non-physical labeling between the physical length and the comoving coordinate.
Due to the rescale of $\bar{a}=a\ee^{\zeta_\uL}$, a specific physical length in the local patch is differently labeled to the comoving coordinate by the factor $\ee^{-\zeta_\uL}$ from the background universe.
It leads to an apparent modulation on the two-point function $\xi_\zeta(x)=\braket{\zeta(\bfy)\zeta(\bfy+\bfx)}$ as
\bae{
    \bar{\xi}_\zeta(x)=\xi_\zeta\pqty{x\ee^{\zeta_\uL}}\simeq\pqty{1+\zeta_\uL x_i\partial_{x_i}}\xi_\zeta(x), 
}
at linear order in $\zeta_\uL$, where $\bar{\xi}_\zeta(x)$ denotes the (small-scale) two-point function modulated by the long-wavelength perturbation.
In Fourier space, one obtains
\bae{\label{eq: Pzetabar}
    \bar{P}_\zeta(k)\simeq\bqty{1-\zeta_\uL\pqty{\dv{\log P_\zeta(k)}{\log k}+3}}P_\zeta(k),
}
making use of integration by parts and the identity $\partial_{x_i}\pqty{x_i\ee^{-i\bfk\cdot\bfx}}=\partial_{k_i}\pqty{k_i\ee^{-i\bfk\cdot\bfx}}$.
If there is no other bispectrum source, one reproduces the consistency relation~\eqref{eq: Maldacena CR} from this equation.

Is it then possible to generalize this discussion to the non-attractor case?
Beyond the attractor limit, there can be two independent DoF, $\phi$ and $\dot{\phi}$, even in a single-field model, and thus it is expected that the long-wavelength perturbation cannot be necessarily absorbed into the coordinate renormalization any longer.
It is in fact clarified as follows, extending the discussion in Ref.~\cite{Bravo:2020hde}.
Let us first explicitly show the ``renormalization" procedure.
In general, the spacetime metric with a long-wavelength perturbation can be written in the comoving gauge as
\bae{\label{eq: comoving metric}
    \dd{s}^2=-\ee^{2\delta\calN_\uL}\dd{t}^2+a^2(t)\ee^{2\zeta_\uL}\pqty{\dd{\bfx}^2+\bm{\calN}_\uL\dd{t}}^2.
}
The linear constraint reads~\cite{Maldacena:2002vr} 
\bae{
    \delta\calN_\uL(t)=\frac{1}{H}\dot{\zeta}_\uL(t) \qc
    \bm{\calN}_\uL(t,\bfx)=\frac{1}{3}\epsilon_H\bfx\dot{\zeta_\uL}(t),
}
in the long-wavelength limit.
Here $\epsilon_H=-\dot{H}/H^2$ is the first slow-roll parameter.
The inflaton field again has no perturbation: $\phi(t,\bfx)=\phi_0(t)$.
The background dynamics is dictated by 
\bae{
    \pdv[2]{\phi_0}{t}+3H\pdv{\phi_0}{t}+\pdv{}{\phi_0}V(\phi_0)=0
    \qc 3\Mpl^2H^2=\frac{1}{2}\pqty{\pdv{\phi_0}{t}}^2+V.
}
The interesting point is that, in this long-wavelength limit, the local dynamics can be understood like as 
a background one without perturbations.
That is, one can take a coordinate transformation
\bae{\label{ct}
    t\to\bar{t}=t+\xi^0(t) \qc
    \bfx\to\bar{\bfx}=\ee^{\beta(t)/3}\bfx,
}
such that the metric shows a ``background-like" form
\bae{
    \dd{s}^2=-\dd{\bar{t}}^2+\bar{a}^2\dd{\bar{\bfx}}^2 \qc
    \bar{a}(\bar{t})=a(t)\ee^{\alpha(t)},
}
and the transformed inflaton $\bar{\phi}(\bar{t})=\phi_0(t)$ follows ``background-like" equations of motion (EoM)
\bae{\label{eq: local EoM}
    \pdv[2]{\bar{\phi}}{\bar{t}}+3\bar{H}\pdv{\bar{\phi}}{\bar{t}}+\pdv{}{\bar{\phi}}V(\bar{\phi})=0 \qc
    3\Mpl^2\bar{H}^2=\frac{1}{2}\pqty{\pdv{\bar{\phi}}{\bar{t}}}^2+V,
}
with $\bar{H}=(\partial_{\bar{t}}\bar{a})/\bar{a}$.
Specifically one can take parameters as
\bae{\label{eq: xi alpha beta}
    \dot{\xi}^0=\frac{1}{H}\dot{\zeta}_\uL \qc
    \alpha+\frac{1}{3}\beta=\zeta_\uL \qc
    \dot{\beta}=\epsilon_H\dot{\zeta}_\uL,
}
and the above speculation is confirmed with use of the EoM for the curvature perturbation on a superhorizon scale:
\bae{\label{eq: zetaL EoM}
    \dv{}{t}\pqty{\epsilon_Ha^3\dot{\zeta}_\uL}=0.
}

However it does not necessarily mean that the long-wavelength perturbation does not affect the local physics.
As EoM~\eqref{eq: local EoM} are second order in $\bar{\phi}$ and first order in $\bar{a}$ (through $\partial_{\bar{t}}\bar{a}=\bar{a}\bar{H}$),
their solution is labeled by three independent DoF $\pqty{\bar{\phi}(\bar{t}),\bar{\phi}^\prime(\bar{t}),\bar{a}(\bar{t})}$. Therefore, even if $\bar{\phi}(\bar{t})$ is fixed to the background value $\phi_0(t)$ and the scale factor is in itself irrelevant to physics, the local observer can distinguish the long-wavelength mode from the background variable via the time derivative $\bar{\phi}^\prime(\bar{t})$, which was not included in Ref.~\cite{Bravo:2020hde}. In fact the time derivative in the local patch is different from the background one by
\bae{
    \bar{\phi}^\prime(\bar{t})=\pqty{\pdv{\bar{t}}{t}}^{-1}\partial_t\bar{\phi}(\bar{t})=\pqty{1+\dot{\xi}^0}^{-1}\dot{\phi}_0,
}
even after the ``renormalization" of the long-wavelength mode.

Such a ``renormalization" is nevertheless useful in order to evaluate the bispectrum in the squeezed limit.
The above discussion still holds even if one includes the short-wavelength perturbation $\zeta_\uS$ in Eq.~\eqref{eq: comoving metric}.
Under the coordinate transformation~(\ref{ct}), 
the short-wavelength curvature perturbation transforms as a scalar~\cite{Bravo:2020hde},\footnote{Notice that ${\bar t}=\text{const.}$ hypersurface and the comoving hypersurface approximately coincide only on short scales. }
\bae{\label{zetas-bar}
    \zeta_\uS(t,\bfx)={\bar \zeta} (\bar{t},\bar{\bfx}).
}
Therefore the short-wavelength mode accompanied by the long-wavelength perturbation is equivalent to the short-wavelength solution of the Mukhanov-Sasaki equation on the transformed background $\pqty{\bar{\phi}(\bar{t}),\bar{\phi}^\prime(\bar{t}),\bar{a}(\bar{t})}$ \emph{without} the long-wavelength mode.
As $\bar{\phi}(\bar{t})$ is fixed to $\phi_0(t)$, we label such a solution by $\pqty{\bar{\phi}^\prime(\bar{t}),\bar{a}(\bar{t})}$ as ${\bar \zeta} (\bar{t},\bar{\bfx})=\zeta [\bar{t},\bar{\bfx}\mid\bar{\phi}^\prime(\bar{t}),\bar{a}(\bar{t}) ]$ to express this fact. Then, from Eq.~(\ref{zetas-bar}), we have
\bae{\label{eq: zetaS by general solution}
    \zeta_\uS(t,\bfx)=\zeta\bqty{\bar{t},\bar{\bfx}\mid\bar{\phi}^\prime(\bar{t}),\bar{a}(\bar{t})}.
}
Inclusion of $\bar{\phi}^\prime$-dependence is a new point of this paper.
At linear order in $\zeta_\uL$, the explicit form of barred variables leads to
\bae{
    \zeta_\uS(t,\bfx)\simeq\zeta\left[t+\xi^0,\pqty{1+\frac{\beta}{3}}\bfx\relmiddle{|}\pqty{1-\dot{\xi}^0-\frac{\ddot{\phi}_0}{\dot{\phi}_0}\xi^0}\dot{\phi}_0(t+\xi^0),\pqty{1+\alpha-H\xi^0}a(t+\xi^0)\right].
}
We here note its dependence on the scale factor. The scale factor appears in EoM only through the combination $a^{-1}\partial_\bfx$,\footnote{Note that the Hubble parameter is here understood as a function of $\phi$ and $\dot{\phi}$ rather than the time derivative of the scale factor.}
so that the solution satisfies a general relation $\zeta[t,\bfx\mid\dot{\phi},a]=\zeta[t,C\bfx\mid\dot{\phi},C^{-1}a]$ for a constant $C$.
Therefore, neglecting the time dependence of $(1+\alpha-H\xi^0)$,\footnote{In fact its time derivative $\dot{\alpha}+\epsilon_HH^2\xi^0-H\dot{\xi}^0$ is always 
slow-roll suppressed at most via Eqs.~\eqref{eq: xi alpha beta} and \eqref{eq: zetaL EoM}. Its negligibility is confirmed in a concrete example shown in Fig.~\ref{fig: fNLvsdelta}.}
one goes further
\bae{\label{eq: zetaS by the general solution}
    \zeta_\uS(t,\bfx)\simeq\zeta\left[t+\xi^0,\pqty{1+\zeta_\uL-H\xi^0}\bfx\relmiddle{|}\pqty{1-\dot{\xi}^0-\frac{\ddot{\phi}_0}{\dot{\phi}_0}\xi^0}\dot{\phi}_0(t+\xi^0),a(t+\xi^0)\right].
}
Therefore the short-wavelength mode is equivalent to the curvature perturbation solution on the modified background momentum $\dot{\phi}=\pqty{1-\dot{\xi}^0-\frac{\ddot{\phi}_0}{\dot{\phi}_0}\xi^0}\dot{\phi}_0$ at the shifted spacetime point $\pqty{t+\xi^0,(1+\zeta_\uL-H\xi^0)\bfx}$.
In terms of the Fourier-space two-point function, the long-wavelength modulation on the short mode is expressed by
\bae{\label{eq: modulated 2pt}
    \braket{\zeta_\uS(\bfk_1)\zeta_\uS(\bfk_2)}&=\braket{\zeta(\bfk_1)\zeta_\uS(\bfk_2)}+\xi^0(\bfk_\uL)\dot{P}_\zeta[t,k_\uS\mid\dot{\phi}_0]-\pqty{\zeta_\uL(\bfk_\uL)-H\xi^0(\bfk_\uL)}(\ns-1)P_\zeta[t,k_\uS\mid\dot{\phi}_0] \nonumber \\
    &\quad -\pqty{\dot{\phi}_0\dot{\xi}^0(\bfk_\uL)+\ddot{\phi}_0\xi^0(\bfk_\uL)}\partial_{\dot{\phi}}P_\zeta[t,k_\uS\mid\dot{\phi}]_{\dot{\phi}=\dot{\phi}_0}, 
}
where the first term denotes the fiducial two-point function without long-wavelength perturbations, and the ensemble average is performed for the short-wavelength modes under the fixed long-wavelength mode $\xi^0$.
Here we restore the spatial dependence of the long-wavelength mode with $\bfk_1+\bfk_2+\bfk_\uL=\mathbf{0}$ and $k_\uS=k_1\gg k_\uL$.
The squeezed-limit bispectrum is given by its correlation with the long-wavelength mode:
\bae{
    \braket{\zeta_\uL(\bfk_3)\braket{\zeta_\uS\zeta_\uS}(\bfk_1,\bfk_2)}=\lim_{k_3\to0}(2\pi)^3\delta^{(3)}(\bfk_1+\bfk_2+\bfk_3)B_\zeta(k_1,k_2,k_3).
}
Thus one obtains the generalized single-field soft theorem in a formal expression as
\bae{\label{eq:gsfst} 
    \lim_{k_3\to0}B_\zeta(t;k_1,k_2,k_3)&=(1-\ns)P_\zeta(t,k_\uL)P_\zeta(t,k_\uS) \nonumber \\
    &\quad+\pqty{\int^t\frac{\partial_{t^\prime}P_\zeta(t,t^\prime;k_\uL)}{H}\dd{t^\prime}}\bqty{\dot{P}_\zeta(t,k_\uS)+H(\ns-1)P_\zeta(t,k_\uS)} \nonumber \\
    &\quad-\bqty{\frac{\dot{\phi}_0}{2H}\dot{P}_\zeta(t,k_\uL)+\ddot{\phi}_0\int^t\frac{\partial_{t^\prime}P_\zeta(t,t^\prime;k_\uL)}{H}\dd{t^\prime}}\partial_{\dot{\phi}}P_\zeta[t,k_\uS\mid\dot{\phi}]_{\dot{\phi}=\dot{\phi}_0},
}
where $P_\zeta(t,t^\prime;k)$ denotes the unequal time two-point function: $\braket{\zeta(t,\bfk)\zeta(t^\prime,\bfk^\prime)}=(2\pi)^3\delta^{(3)}(\bfk+\bfk^\prime)P_\zeta(t,t^\prime;k)$.
The lower limit of the time integration of $\dot{\xi}^0$ should be chosen so that the resultant $\xi^0$ properly satisfies its definition $\bar{\phi}(t+\xi^0)=\phi_0(t)$.

The last term represents the $\dot{\phi}$ correction to the generalized ``consistency relation" proposed in Refs.~\cite{Bravo:2017wyw,Bravo:2017gct,Bravo:2020hde} (see also Ref.~\cite{Finelli:2017fml} deriving the same formula but with a restriction to the exact shift symmetry).
In a pure attractor case, the time-independence of $\zeta_\uL$ recovers Maldacena's consistency relation~\eqref{eq: Maldacena CR} from this formula.
In the USR limit, $|V^\prime|\ll|3H\dot{\phi}_0|$, as an extreme example of non-attractor models,
the curvature perturbation grows as $\zeta\propto a^3$ and
one finds that the last term again vanishes, making use of the background EoM $\ddot{\phi}_0+3H\dot{\phi}_0=0$.
It could be understood to reflect the fact that there is effectively only one DoF $\dot{\phi}$ during the exact USR phase and the effect of such a single DoF can be expressed by the spacetime coordinate renormalization.
Making use of the time-dependence $\dot{P}_\zeta=6HP_\zeta$ and the scale-invariance $\ns-1=0$ of curvature perturbations, one recovers the well-known result in the USR limit, $\fNL=5/2$, in terms of the non-linearity parameter
\bae{
    \fNL=\frac{5}{12}\frac{B_\zeta(k_\uL,k_\uS,k_\uS)}{P_\zeta(k_\uL)P_\zeta(k_\uS)}.
}
These results in the two extreme cases are consistent with the previous study~\cite{Bravo:2020hde}.
However the last correction does not vanish in general and can give a non-zero contribution to the squeezed bispectrum.
In particular, once inflation experiences an USR phase, even if it is followed by a SR phase and $\dot{\phi}$ quickly converges to the attractor value, the tiny difference in $\dot{\phi}$ keeps retaining the information of perturbations during USR and yielding a non-vanishing bispectrum well after the transition as we see in the rest of the paper.

We here fix the field value $\phi$ and express the bispectrum correction in terms of the momentum-dependence of the power spectrum. We note that it is not necessary: one can instead fix the momentum and find the correction as the $\phi$-dependence of the power spectrum to see the difference in the phase-space trajectory. In Appendix~\ref{sec: uniform-dotphi}, we concretely discuss this equality and successfully obtain the same formula~\eqref{eq: Suyama formula} we will derive in the next section.

\section{Bispectrum in the transition from ultra slow-roll to slow-roll}
\label{Sec:Bis}

We here focus on the models starting from an USR phase and followed by a SR phase.
Making use of the modulated two-point function~\eqref{eq: modulated 2pt}, we derive a general formula for the squeezed bispectrum well after the transition from USR to SR.
Because perturbations exiting the horizon during USR are known to be scale-invariant $\ns-1=0$ and also the curvature perturbation ceases its time evolution $\dot{P}_\zeta\to0$ (in the deep SR phase), all we need is to vary the momentum and find $P_\zeta$'s dependence on it.
The schematic figure~\ref{fig: USR2SR} summarizes our setup.

\bfe{width=0.7\hsize}{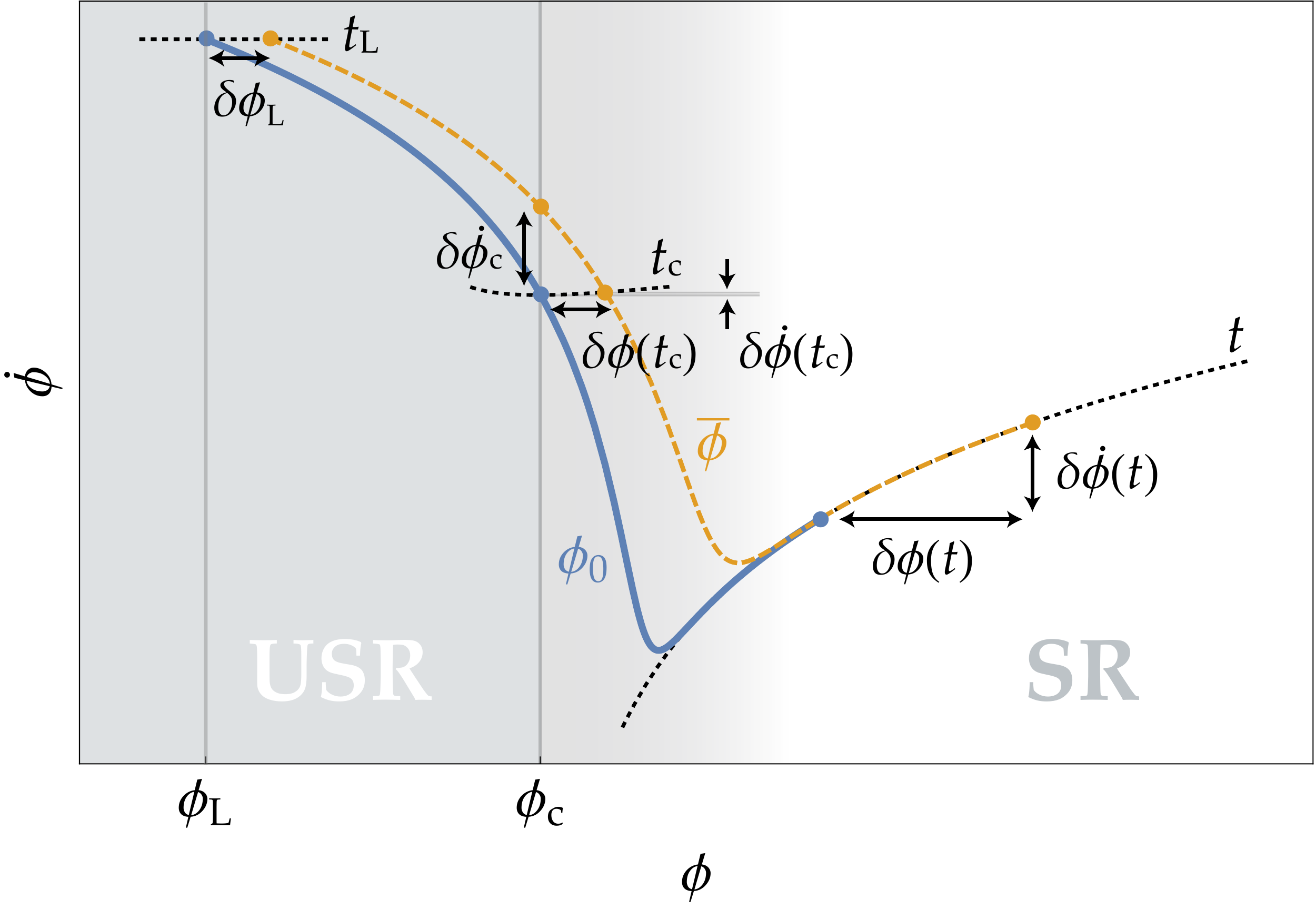}{The schematic picture of our notation and setup to study the transition from the ultra slow-roll phase (gray region) to the slow-roll phase (white region).
The inflaton $\phi$ evolves from left to right without loss of generality. The inflaton dynamics is assumed to be well approximated by the USR solution before some point $\phi_\uc$. After $\phi_\uc$, the dynamics gradually or rapidly deviates from the USR one and it finally reaches the SR phase, where we evaluate the squeezed bispectrum at the time $t$.
The blue line illustrates the unperturbed background trajectory $\phi_0$, while the orange dashed line represents the perturbed (shifted) trajectory $\bar{\phi}$~\eqref{eq: def of phibar}. 
Black dotted lines are the equal-time hypersurfaces.
At the initial time $t_\uL$, the long-wavelength perturbation should be represented only by the field difference $\delta\phi_\uL$ with the equated momenta: $\bar{\phi}^\prime(t_\uL)=\dot{\phi}_0(t_\uL)$~\eqref{eq: def of delta phi}.
This field difference nevertheless yields the momentum perturbation $\delta\dot{\phi}_\uc$ at $\phi=\phi_\uc$~\eqref{eq: sol of delta phidotc}, which characterizes the later perturbation. After $\phi=\phi_\uc$, we allow a general potential form as long as it converges to a SR-type one.
Converting $\delta\dot{\phi}_\uc$ to the initial condition of the perturbation at $t=t_\uc$, $\delta\phi(t_\uc)$ and $\delta\dot{\phi}(t_\uc)$~\eqref{eq: perturbation IC}, we consider the solution of the perturbed EoM~\eqref{eq: ptb EoM with F} along such a potential.
}{fig: USR2SR}

Let us first assume the situation that the USR condition is well satisfied before the inflaton reaches some value $\phi_\uc$, while the potential can start to deviate from the USR one (or even it can suddenly change) at that point.
Then the chain rule leads to
\bae{\label{eq: dPzddotphi}
    \pdv{P_\zeta}{\dot{\phi}}=\pqty{\pdv{\dot{\phi}}{\dot{\phi}_\uc}}^{-1}\pdv{P_\zeta}{\dot{\phi}_\uc},
}
where $\dot{\phi}_\uc$ is the momentum at $\phi=\phi_\uc$.
It is also useful to define the power of $P_\zeta$ in $\dot{\phi}_\uc$ around the background trajectory by
\bae{\label{eq: power-law assumption}
    n\coloneqq\eval{-\pdv{\log P_\zeta}{\log\dot{\phi}_\uc}}_{\phi=\phi_0},
}
so that Eq.~\eqref{eq: dPzddotphi} reads
\bae{
    \pdv{P_\zeta}{\dot{\phi}}=-n\pqty{\pdv{\dot{\phi}}{\dot{\phi}_\uc}}^{-1}\frac{P_\zeta}{\dot{\phi}_\uc}.
}
Then the remained non-trivial part is the $\dot{\phi}_\uc$ dependence of the momentum value $\dot{\phi}$ at the evaluation time.

We then specifically consider the shifted trajectory $\bar{\phi}(t)$ from the background solution $\phi_0(t)$. How should this $\bar{\phi}(t)$ be defined?
To utilize the formula~\eqref{eq: zetaS by the general solution}, one finds that the shifted trajectory $\bar{\phi}(t)$ and the time shift $\xi^0$ should satisfy
\bae{\label{eq: def of phibar}
    \bar{\phi}(t+\xi^0)=\phi_0(t) \qc
    \bar{\phi}^\prime(t)=\pqty{1-\dot{\xi}^0-\frac{\ddot{\phi}_0}{\dot{\phi}_0}\xi^0}\dot{\phi}_0(t).
}
In particular, the momentum shift vanishes, $\pqty{1-\dot{\xi}^0-\frac{\ddot{\phi}_0}{\dot{\phi}_0}\xi^0}\dot{\phi}_0=\dot{\phi}_0$, during USR as we mentioned in the second-to-last paragraph of the previous section.
Therefore, at some initial time $t_\uL$ in the USR phase, we only shift the field value and keep the momentum intact:
\bae{\label{eq: def of delta phi}
    \bar{\phi}(t_\uL)=\phi_0(t_\uL)+\delta\phi_\uL \qc
    \bar{\phi}^\prime(t_\uL)=\dot{\phi}_0(t_\uL).
}
From this initial condition, we need to derive the $\dot{\phi}_\uc$ modulation
\bae{
    \delta\dot{\phi}_\uc\coloneqq\bar{\phi}^\prime(\bar{t}_\uc)-\dot{\phi}_0(t_\uc), \quad
    \text{with $\bar{\phi}(\bar{t}_\uc)=\phi_\uc$},
}
and the perturbation at the evaluation time
\bae{
    \delta\phi(t)\coloneqq\bar{\phi}(t)-\phi_0(t) \qc
    \delta\dot{\phi}(t)\coloneqq\bar{\phi}^\prime(t)-\dot{\phi}_0(t),
}
so that the chain-rule coefficient is given by
\bae{
    \pqty{\pdv{\dot{\phi}(t)}{\dot{\phi}_\uc}}^{-1}=\eval{\frac{\delta\dot{\phi}_\uc}{\delta\dot{\phi}(t)}}_{\delta\phi_\uL\to0}.
}

The numerator $\delta\dot{\phi}_\uc$ can be easily obtained. With use of the USR EoM $\ddot{\phi}+3H_\uc\dot{\phi}=0$ with a constant Hubble parameter $H_\uc$, one finds
\bae{\label{eq: sol of delta phidotc}
    \delta\dot{\phi}_\uc=3H_\uc\delta\phi_\uL.
}
The time shift at linear order,
\bae{
    \delta t_\uc\coloneqq\bar{t}_\uc-t_\uc\simeq-\frac{\delta\phi_\uL}{\dot{\phi}_\uc},
}
is also useful.
We then divide the shifted solution into the USR and SR ones as
\bae{
    \bar{\phi}(t)=\bce{
        \bar{\phi}_\USR(t), & \text{for $t<\bar{t}_\uc$}, \\
        \bar{\phi}_\SR(t), & \text{for $t\geq \bar{t}_\uc$}.
    }
}
The initial condition for the perturbation can be set at $t_\uc$ as follows at linear order in $\delta\phi_\uL$.\footnote{Depending on $\delta\phi_\uL$, the shifted solution $\bar{\phi}$ can be still in the USR phase at $t_\uc$. To consider only the SR phase, we however extrapolate the SR solution $\bar{\phi}_\SR$ to the time $t_\uc$ and set the perturbation initial condition with use of $\bar{\phi}_\SR$ in that case.}
\bae{\label{eq: perturbation IC}
    \bce{
        \dps
        \delta\phi(t_\uc)\coloneqq\bar{\phi}_\SR(t_\uc)-\phi_0(t_\uc)=\bar{\phi}_\SR(\bar{t}_\uc-\delta t_\uc)-\phi_\uc\simeq-\bar{\phi}_\SR^\prime(\bar{t}_\uc)\delta t_\uc\simeq-\dot{\phi}_\uc\delta t_\uc\simeq\delta\phi_\uL, \\
        \dps
        \delta\dot{\phi}(t_\uc)\coloneqq\bar{\phi}_\SR^\prime(t_\uc)-\dot{\phi}_0(t_\uc)=\bar{\phi}_\SR^\prime(\bar{t}_\uc-\delta t_\uc)-\dot{\phi}_\uc\simeq\delta\dot{\phi}_\uc-\bar{\phi}_\SR^{\prime\prime}(\bar{t}_\uc)\delta t_\uc\simeq-\frac{V^\prime_\SR(\phi_\uc)}{\dot{\phi}_\uc}\delta\phi_\uL.
    }
}

Once we express the background EoM in the SR phase by
\bae{\label{eq: background EoM with F}
    \ddot{\phi}=-3H(\phi,\dot{\phi})\dot{\phi}-V(\phi)\eqqcolon F(\phi,\dot{\phi}),
}
the perturbation EoM can be written as
\bae{\label{eq: ptb EoM with F}
    \delta\ddot{\phi}=F_{\dot{\phi}}\delta\dot{\phi}+F_\phi\delta\phi,
}
where $F_{\dot{\phi}}=\partial_{\dot{\phi}}F$ and $F_\phi=\partial_\phi F$.
On the other hand, the time derivative of the background EoM~\eqref{eq: background EoM with F} reads
\bae{
    \dddot{\phi}=F_{\dot{\phi}}\ddot{\phi}+F_\phi\dot{\phi}.
}
Comparing them, one finds that $u_1=\dot{\phi}_0$ is one solution of the perturbation EoM~\eqref{eq: ptb EoM with F}.
Another independent solution $u_2$ can be found through the Wronskian $W=u_1\dot{u}_2-\dot{u}_1u_2$. The Wronskian satisfies
\bae{
    \dot{W}=F_{\dot{\phi}}W,
}
whose solution is obviously given by $W=\exp\pqty{\int^t_{t_\uc}F_{\dot{\phi}}\dd{t^\prime}}$.
$u_2$ is then formally solved as
\bae{
    u_2(t)=\dot{\phi}_0(t)\int^t_{t_\uc}\frac{W(t^\prime)}{\dot{\phi}_0^2(t^\prime)}\dd{t^\prime}.
}
Noting $u_2(t_\uc)=0$, the perturbation solution with the initial condition~\eqref{eq: perturbation IC} is expressed by
\bae{
    \delta\phi(t)&=\frac{\delta\phi(t_\uc)}{u_1(t_\uc)}\pqty{u_1(t)-\frac{\dot{u}_1(t_\uc)}{\dot{u}_2(t_\uc)}u_2(t)}+\frac{\delta\dot{\phi}(t_\uc)}{\dot{u}_2(t_\uc)}u_2(t) \nonumber \\
    &\simeq\pqty{\frac{\dot{\phi}_0(t)}{\dot{\phi}_\uc}+3H_\uc\dot{\phi}_\uc\dot{\phi}_0(t)\int^t_{t_\uc}\frac{W}{\dot{\phi}_0^2}\dd{t^\prime}}\delta\phi_\uL.
}
Therefore the chain-rule coefficient reads
\bae{\label{eq: ddotphiddotphic}
    \pqty{\pdv{\dot{\phi}(t)}{\dot{\phi}_\uc}}^{-1}=3H_\uc\pqty{\frac{\ddot{\phi}_0(t)}{\dot{\phi}_\uc}+3H_\uc\dot{\phi}_\uc\ddot{\phi}_0(t)\int^t_{t_\uc}\frac{W}{\dot{\phi}_0^2}\dd{t^\prime}+3H_\uc\dot{\phi}_\uc\frac{W(t)}{\dot{\phi}_0(t)}}^{-1}.
}
Given that $F_{\dot{\phi}}=-3H+\cdots$ where dots represent slow-roll corrections, the Wronskian $W=\exp\pqty{\int^t_{t_\uc}F_{\dot{\phi}}\dd{t^\prime}}$ quickly converges to zero for $t\gg t_\uc$. Hence, in the deep SR phase, one has
\bae{\label{eq: Zc}
    \pqty{\pdv{\dot{\phi}}{\dot{\phi}_\uc}}^{-1}\to\frac{3H_\uc\dot{\phi}_\uc}{\ddot{\phi}_0(t)Z_\uc} \qc
    Z_\uc\coloneqq1+3H_\uc\dot{\phi}_\uc^2\int^\infty_{t_\uc}\frac{\exp\pqty{\int^t_{t_\uc}F_{\dot{\phi}}\dd{t^\prime}}}{\dot{\phi}_0^2(t)}\dd{t}.
}
Noting that $\dot{\xi}_0=\dot{\zeta}_\uL/H$ quickly disappears soon after $\phi_\uc$ and thus $\xi^0$ can be approximated by the last value $\zeta_\uL/H_\uc$ in the deep SR phase (correction is suppressed by the first slow-roll parameter), one finally finds a simple formula for the bispectrum correction as
\bae{\label{eq: bispectrum correction}
    -\pqty{\dot{\phi}_0\dot{\xi}^0+\ddot{\phi}_0\xi^0}\partial_{\dot{\phi}}P_\zeta[t,k_\uS\mid\dot{\phi}]_{\dot{\phi}=\dot{\phi}_0}\to\frac{3n}{Z_\uc}\zeta_\uL P_\zeta(k_\uS).
}
Recalling that $\ns-1=0$ and $\dot{P}_\zeta\to0$ in our case, it leads to the non-linearity parameter
\bae{\label{eq: Suyama formula}
    \fNL\to\frac{5n}{4Z_\uc}.
}
Therefore the non-zero squeezed bispectrum remains even after the USR phase unless $n=0$.

\section{Examples}
\label{Sec:Exa}

Let us show several concrete consequences of our formula~\eqref{eq: Suyama formula} in specific transition models in this section.

\subsection{Linear slope with sharp transition}

We first consider a sharp transition from an exactly flat potential to a constant linear slope:
\bae{\label{eq: sharp linear slope}
    V(\phi)=\bce{
        V_0, & \text{for $\phi<\phi_\uc$}, \\
        \dps
        V_0\pqty{1-\sqrt{2\epsilon_0}\frac{\phi-\phi_\uc}{\Mpl}},  
        & \text{for $\phi\geq\phi_\uc$}.
    }
}
The Hubble parameter is approximated to be constant, $3\Mpl^2H^2\simeq V_0$.
The constant parameter $\epsilon_0$ corresponds to the first slow-roll parameter $\epsilon_H$ in the deep SR phase.
For this simple potential, one finds an analytic solution for the background as
\bae{
    \phi_0(t)=\bce{
        \dps
        \phi_\uL+\frac{\dot{\phi}_\uL}{3H}\pqty{1-\ee^{-3H(t-t_\uL)}}, & \text{for $\phi_0<\phi_\uc$}, \\[5pt]
        \dps
        \phi_\uc-\frac{1}{3H}\pqty{\pi_0-\dot{\phi}_\uL\ee^{-3H(t_\uc-t_\uL)}}\pqty{1-\ee^{-3H(t-t_\uc)}}+\pi_0(t-t_\uc), & \text{for $\phi_0\geq\phi_\uc$},
    }
}
with the $\phi_\uc$-crossing time
\bae{
    t_\uc=t_\uL-\frac{1}{3H}\log\bqty{1-\frac{3H}{\dot{\phi}_\uL}(\phi_\uc-\phi_\uL)},
}
the terminal velocity
\bae{
    \pi_0=\sqrt{2\epsilon_0}\Mpl H,
}
and the initial condition
\bae{
    \phi_0(t_\uL)=\phi_\uL \qc
    \dot{\phi}_0(t_\uL)=\dot{\phi}_\uL.
}
For a shifted trajectory
\bae{
    \bar{\phi}(t_\uL)=\phi_\uL+\delta\phi_\uL \qc
    \bar{\phi}^\prime(t_\uL)=\dot{\phi}_\uL,
}
its analytic expression is also easily obtained as
\bae{
    \bar{\phi}(t)=\bce{
        \dps
        \phi_\uL+\delta\phi_\uL+\frac{\dot{\phi}_\uL}{3H}\pqty{1-\ee^{-3H(t-t_\uL)}}, & \text{for $\bar{\phi}<\phi_\uc$}, \\[5pt]
        \dps
        \phi_\uc-\frac{1}{3H}\pqty{\pi_0-\dot{\phi}_\uL\ee^{-3H(\bar{t}_\uc-t_\uL)}}\pqty{1-\ee^{-3H(t-\bar{t}_\uc)}}+\pi_0(t-\bar{t}_\uc), & \text{for $\bar{\phi}\geq\phi_\uc$},
    }
}
with
\bae{
    \bar{t}_\uc=t_\uL-\frac{1}{3H}\log\bqty{1-\frac{3H}{\dot{\phi}_\uL}(\phi_\uc-\phi_\uL-\delta\phi_\uL)}.
}
From the condition $\bar{\phi}(t+\xi^0)=\phi_0(t)$, one further finds the solution for $\xi^0$ at linear order in $\delta\phi_\uL$ as
\bae{
    \xi^0(t)\simeq\bce{
        \dps
        -\frac{\delta\phi_\uL}{\dot{\phi}_\uL}\ee^{3H(t-t_\uL)}, & \text{for $t<t_\uc$} \\[5pt]
        \dps
        -\frac{\dot{\phi}_\uc+\pi_0\pqty{1-\ee^{-3H(t-t_\uc)}}}{\dot{\phi}_\uc\pqty{\dot{\phi}_\uc\ee^{-3H(t-t_\uc)}+\pi_0\pqty{1-\ee^{-3H(t-t_\uc)}}}}\delta\phi_\uL, & \text{for $t\geq t_\uc$}.
    }
}
As it is directly related with the curvature perturbation by $\zeta_\uL=H\xi^0$ in the constant $H$ approximation, this expression shows the exponential growth of $\zeta_\uL$ during the USR phase as well as its asymptotic value in the deep SR phase $\zeta_\uL\to-H\frac{\delta\phi_\uL}{\dot{\phi}_\uc}$ in the large transition limit $\pi_0\gg \dot{\phi}_\uc$.
Furthermore, this relation is applicable also for short-wavelength modes with a replacement $\delta\phi_\uL\to\delta\phi_\uS$. One then finds
\bae{
    P_\zeta(k_\uS)\to\frac{H^2}{\dot{\phi}_\uc^2}P_{\delta\phi}(k_\uS),
}
which corresponds to the power of $n=2$ defined in Eq.~\eqref{eq: power-law assumption}.
$Z_\uc$'s integration can be also explicitly done to obtain
\bae{
    Z_\uc=1+\frac{\dot{\phi}_\uc}{\pi_0}\simeq1.
}
Therefore our formula~\eqref{eq: Suyama formula} successfully reproduces the well-studied result by the rigorous in-in formalism in the large transition limit~\cite{Cai:2017bxr,Passaglia:2018ixg},
\bae{
    \fNL\to\frac{5}{2}.
}

We mention that the last term in Eq.~\eqref{eq: ddotphiddotphic} and the first term in Eq.~\eqref{eq: bispectrum correction} can be actually non-negligible in the linear slope case (vanishing $V^{\prime\prime}$) if the transition is not sufficiently large because $\ddot{\phi}_0$ keeps decaying as well as $W(t)$ and $\dot{\phi}_0\dot{\xi}^0$ in this case.
If one keeps all relevant terms without any large transition approximation, our formula is modified as
\bae{
    \fNL\to\frac{1}{1-\delta^2}\frac{5n}{4Z_\uc},
}
with
\bae{
    n\to\frac{2}{1+\delta} \qc
    Z_c\to\frac{1}{1-\delta},
}
where $\delta=\dot{\phi}_\uc/\pi_0$.
The non-linaerity parameter then reads
\bae{\label{eq: Cai formula}
    \fNL\to\frac{5}{2}\frac{1}{(1+\delta)^2},
}
which is exactly equivalent to the formula~(2.45) in Ref.~\cite{Cai:2017bxr} with $\eta_V=0$.\footnote{Our $\delta$ is equivalent to $-6/h$ of Ref.~\cite{Cai:2017bxr}.}

\subsection{Linear slope with smooth transition}

The previous linear slope~\eqref{eq: sharp linear slope} is somewhat unphysical because its first derivative is discontinuous at $\phi_\uc$. In this subsection, we smoothly connect the flat potential and the linear slope, referring to the prescription in Ref.~\cite{Passaglia:2018ixg}.
That is, we consider the potential
\bae{
    V(\phi)=V_0\bqty{1-\sqrt{2\epsilon_0}\frac{\phi-\phi_\ut+d_\phi\log\pqty{2\cosh\frac{\phi-\phi_\ut}{d_\phi}}}{2\Mpl}},
}
so that its first derivative is continuous at $\phi_\ut$ as
\bae{
    V^\prime(\phi)=-\frac{\sqrt{2\epsilon_0}V_0}{2\Mpl}\pqty{1+\tanh\frac{\phi-\phi_\ut}{d_\phi}},
}
with the smoothness parameter $d_\phi$.
Note that here $\phi_\uc$ is not equivalent to $\phi_\ut$ but we rather take $\phi_\uc$ much before $\phi_\ut$ so that the inflaton dynamics is well approximated by the USR solution until $\phi_\uc$ to utilize our formula~\eqref{eq: Suyama formula}. In this case, the $\dot{\phi}_\uc$-dependence $n$ is not necessarily equal to two and $Z_\uc$ is not estimated as unity.
Hence we numerically evaluate $n$ and $Z_\uc$: the power spectrum $P_\zeta$ is calculated with use of the $\delta N$ formalism along various shifted background trajectories to find its $\dot{\phi}_\uc$ ($\dot{\phi}$ at $\phi=\phi_\uc$) dependence~\eqref{eq: power-law assumption}, while $Z_\uc$~\eqref{eq: Zc} is obtained by the numerical integration along the unperturbed trajectory.

On the other hand, Passaglia, Hu, and Motohashi (PHM) in Ref.~\cite{Passaglia:2018ixg} modeled this transition by the effective $\delta$ parameter characterized by the largeness and fastness of the transition.\footnote{Our $\delta$ is equivalent to $1/h$ of Ref.~\cite{Passaglia:2018ixg}.}
They determined the transition regime as follows. First the end of transition is defined by
\bae{
    \phi^\uf_\ut\coloneqq\phi_\ut+2d_\phi.
}
On the other hand, the beginning of the transition $\phi^\ui_\ut$ is not simply given by $\phi_\ut-2d_\phi$ but instead determined by the deviation from the USR solution as
\bae{
    1-\eval{\frac{\dot{\phi}_\USR}{\dot{\phi}}}_{\phi^\ui_\ut}\coloneqq0.05\pqty{1-\eval{\frac{\dot{\phi}_\USR}{\dot{\phi}}}_{\phi^\uf_\ut}},
}
where $\phi_\USR$ is the USR solution
\bae{
    \dot{\phi}_\USR(t)=\dot{\phi}_\uL\ee^{-3N(t)},
}
with the e-folding number $N(t)=\int^t_{t_\uL}H\dd{t^\prime}$.
The fastness of the transition is characterized by the e-folding number during the transition regime:
\bae{\label{eq: dN}
    d_N\coloneqq \eval{N}_{\phi^\uf_\ut}-\eval{N}_{\phi^\ui_\ut}.
}
The largeness of the transition is defined by
\bae{\label{eq: delta}
    \delta\coloneqq\sqrt{\frac{\eval{\epsilon_V}_{\phi^\uf_\ut}}{\eval{\epsilon_H}_{\phi^\ui_\ut}}},
}
with
\bae{
    \epsilon_V=\frac{\Mpl^2}{2}\pqty{\frac{V^\prime}{V}}^2 \qc
    \epsilon_H=-\frac{\dot{H}}{H^2}=\frac{\dot{\phi}^2}{2\Mpl^2H^2}.
}
PHM then showed that the non-linearity parameter computed in the full in-in approach is well approximated by the same formula~\eqref{eq: Cai formula} for the sharp transition but with the effective $\delta$ parameter as
\bae{\label{eq: PHM model}
    \fNL\simeq\frac{5}{2}\frac{1}{(1+\delta_\eff)^2} \qc
    \delta_\eff=\bqty{\pqty{\frac{d_N}{1.5}}^3+\delta^3}^{1/3}.
}
In Fig.~\ref{fig: fNLvsdelta}, we compare our formula~\eqref{eq: Suyama formula} with this PHM model, varying the fastness $d_N$ and the largeness $\delta$.
We also checked that our formula is well consistent with the full $\delta N$ computation.

\bfe{width=0.7\hsize}{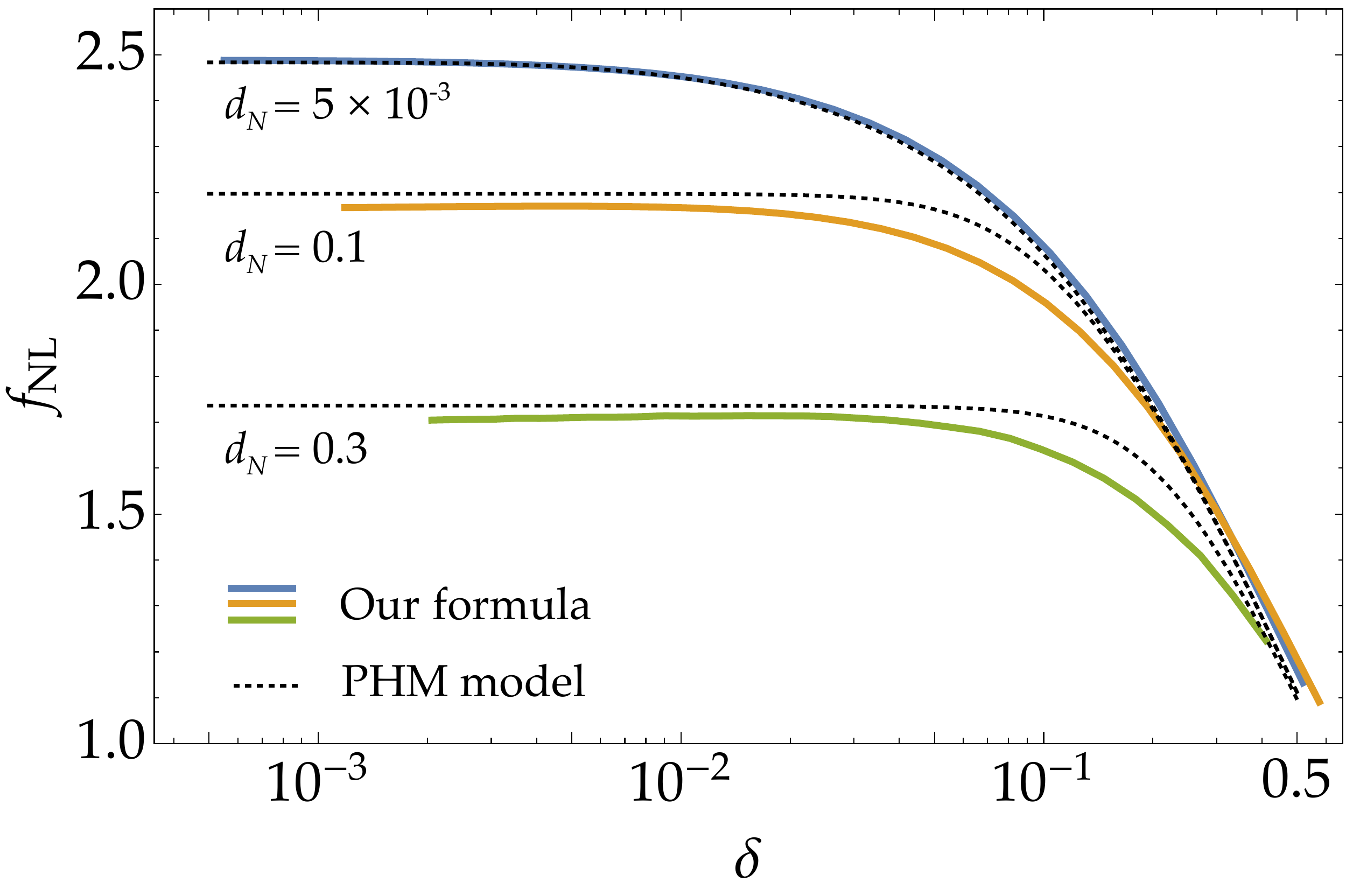}{The $\fNL$ prediction in our formula~\eqref{eq: Suyama formula} (thick lines) and PHM model~\eqref{eq: PHM model} (dotted lines), varying the fastness $d_N$~\eqref{eq: dN} (blue for $d_N=5\times10^{-3}$, orange for $d_N=0.1$, and green for $d_N=0.3$) and the largeness $\delta$~\eqref{eq: delta} of the transition. It is also checked that our formula is well consistent with the full $\delta N$ computation though we do not explicitly plot their results because they cannot be distinguished from our formula's results in this figure.}{fig: fNLvsdelta}

\subsection{Quadratic slope}

We also consider a quadratic slope with a smooth transition:
\bae{
    V(\phi)=\bce{
        V_0, & \text{for $\phi<\phi_\uc$}, \\
        \dps
        V_0-\frac{1}{2}m^2(\phi-\phi_\uc)^2, & \text{for $\phi\geq\phi_\uc$},
    }
}
with a constant Hubble parameter $3\Mpl^2H^2\simeq V_0$.
The analytic solution for the background reads
\bae{
    \phi_0(t)=\bce{
        \dps
        \phi_\uL+\frac{\dot{\phi}_\uL}{3H}\pqty{1-\ee^{-3H(t-t_\uL)}}, & \text{for $\phi_0<\phi_\uc$}, \\[5pt]
        \dps
        \phi_\uc+\frac{\dot{\phi}_\uL}{C_+-C_-}\ee^{-3H(t_\uc-t_\uL)}\pqty{\ee^{C_+(t-t_\uc)}-\ee^{C_-(t-t_\uc)}}, & \text{for $\phi_0\geq\phi_\uc$},
    }
}
where 
\bae{
    C_\pm=\frac{-3H\pm\sqrt{9H^2+4m^2}}{2} \qc
    t_\uc=t_\uL-\frac{1}{3H}\log\bqty{1-\frac{3H}{\dot{\phi}_\uL}(\phi_\uc-\phi_\uL)}.
}
The shifted trajectory is given by
\bae{
    \bar{\phi}(t)=\bce{
        \dps
        \phi_\uL+\delta\phi_\uL+\frac{\dot{\phi}_\uL}{3H}\pqty{1-\ee^{-3H(t-t_\uL)}}, & \text{for $\bar{\phi}<\phi_\uc$}, \\[5pt]
        \dps
        \phi_\uc+\frac{\dot{\phi}_\uL}{C_+-C_-}\ee^{-3H(\bar{t}_\uc-t_\uL)}\pqty{\ee^{C_+(t-\bar{t}_\uc)}-\ee^{C_-(t-\bar{t}_\uc)}}, & \text{for $\bar{\phi}\geq\phi_\uc$},
    }
}
with 
\bae{
    \bar{t}_\uc=t_\uL-\frac{1}{3H}\log\bqty{1-\frac{3H}{\dot{\phi}_\uL}(\phi_\uc-\phi_\uL-\delta\phi_\uL)}.
}
One finds the time shift $\xi^0$ as 
\bae{
    \xi^0(t)\simeq\bce{
        \dps
        -\frac{\delta\phi_\uL}{\dot{\phi}_\uL}\ee^{3H(t-t_\uL)}, & \text{for $t<t_\uc$}, \\[5pt]
        \dps
        \frac{\delta\phi_\uL}{\dot{\phi}_\uc}\frac{C_+-C_-\ee^{(C_+-C_-)(t-t_\uc)}}{C_--C_+\ee^{(C_+-C_-)(t-t_\uc)}}, & \text{for $t\geq t_\uc$},
    }
}
at linear order in $\delta\phi_\uL$.
With a replacement $\delta\phi_\uL\to\delta\phi_\uS$, it again shows that the short-wavelength curvature perturbation converges to the constant
\bae{
    P_\zeta(k_\uS)\to\frac{C_-^2}{C_+^2}\frac{H^2}{\dot{\phi}_\uc^2}P_{\delta\phi}(k_\uS),
}
implying the power of $n=2$.
$Z_\uc$ is also obtained as
\bae{
    Z_\uc\simeq\frac{9H^2}{m^2},
}
at the leading order in the slow-roll approximation $m^2\ll H^2$.
Therefore our formula reads
\bae{\label{eq: quadratic fNL}
    \fNL\to\frac{5m^2}{18H^2}.
}
One finds that the non-linearity is suppressed by the factor $m^2/9H^2$ in the slow-roll limit $m^2\ll H^2$ in this case, compared to the sharp transition case.
Note that $\eta_V=-m^2/3H^2$ in our case and this result is consistent with the formula~(3.36) derived in Ref.~\cite{Atal:2018neu} in the slow-roll limit $|\eta_V|\ll1$.

One can generalize the above linear and quadratic slope discussions to the following mixed slope:
\bae{
    V(\phi)=\bce{
        \dps
        V_0, & \text{for $\phi<\phi_\uc$}, \\
        \dps
        V_0\pqty{1-\sqrt{2\epsilon_0}\frac{\phi-\phi_\uc}{\Mpl}+\frac{\eta_0}{2}\frac{(\phi-\phi_\uc)^2}{\Mpl^2}}, & \text{for $\phi\geq\phi_\uc$},
    }
}
with constant parameters $\epsilon_0$ and $\eta_0$.
Though we do not explicitly show its tedious expression, one finds the analytic solution even for this potential.
In the deep SR limit, the final result is obtained as
\bae{
    \fNL\to\frac{5(3-s)(9-12\delta\eta_0-(s-2\delta\eta_0)^2)}{2(3+s)(3-s+2\delta\eta_0)^2} \qc
    s=\sqrt{9-12\eta_0}.
}
This result is consistent with the formula~(2.55) in Ref.~\cite{Cai:2017bxr} derived with use of the $\delta N$ formula up to $\calO(\eta_0^2)$ corrections.
Their difference mainly comes from the fact that they use the approximated solution~(2.49) for $\phi$ while we employ the exact solution.
Our result consistently reproduces Eq.~\eqref{eq: Cai formula} in the linear slope limit $\eta_0\to0$ and Eq.~\eqref{eq: quadratic fNL} in the smooth transition limit $\delta\to\infty$.

\section{Conclusions and discussion}
\label{Sec:Conclusions}

In this paper, we have revisited the soft limit of the correlation functions of primordial curvature perturbations in non-attractor inflation models with the standard Bunch-Davies initial condition because there are some discrepancies among the literature. A non-attractor (USR) phase needs to be followed by an attractor (SR) phase because, otherwise, USR inflation cannot solve the horizon and flatness problems in general and cannot fit the observed density perturbations. Then, the question is whether Maldacena's consistency relation holds true even for a long wavelength mode exiting the horizon during an USR phase, or not.
Some explicit computations~\cite{Cai:2017bxr,Passaglia:2018ixg} show that it can violate, while the exact USR inflation, a representative of non-attractor models, can be understood by a generalized consistency relation (only during the USR phase though)~\cite{Bravo:2017gct,Bravo:2017wyw,Bravo:2020hde}.

In order to address this question, we have extended the previous formula by including the momentum dependence of the inflaton, which is necessary to deal with a transition phase from an USR one to a SR one adequately. It is true that single degree of freedom (of phase space) is enough to describe the dynamics of an USR or a SR phase respectively, but two degrees of freedom are necessary to describe the transition (and its later dynamics) properly. Then, we have obtained the generalized single-field soft theorem~\eqref{eq:gsfst}, which can deal with the effect of a transition phase as well as an USR phase and/or a SR phase itself. This new formula clearly shows that Maldacena's consistency relation can be violated in general for long-wavelength perturbations exiting the horizon during a non-attractor (USR) phase.

Based on this new formula, we also discussed some examples of non-attractor (USR) inflation models followed by attractor (SR) phases with several types of transitions. Our formula successfully reproduces the results based on the $\delta N$ formula and also those obtained in the existing literature.

Let us comment on the so-called \emph{local observer effect}. The essence of this program is that the comoving-coordinate correlators are contaminated by the long-wavelength metric perturbations, so that the correlators measured by the physical time and length would be helpful to be compared with real observational data.
After the universe converges to an attractor behavior, the only modulation by long-wavelength perturbations is a scale shift due to the fluctuated scale factor $\bar{a}=a\ee^{\zeta_\uL}$ as discussed at the beginning of Sec.~\ref{Sec:Con}. The bispectrum is then corrected by $-(1-\ns)P_\zeta(k_\uL)P_\zeta(k_\uS)$ for local observers. Therefore, in case of an attractor (SR) single-field inflation where the (comoving) bispectrum is given by Maldacena's consistency relation~\eqref{eq: Maldacena CR}, this local observer effect exactly cancels the squeezed limit of the (physical) bispectrum except the secondary effects.
It can be also understood in terms of the backward e-foldings~\cite{Tada:2016pmk}. As the end of inflation gives a fixed physical measure via the Hubble scale $H^{-1}$, the backward e-folding number from the end of inflation labels the physical scale up to the time variation of $H$.
In the attractor and single-field case, an equal backward e-folds hypersurface obviously defines one phase-space point. Thus there is no difference in each physically rescaled local patch in this case.

In the USR limit, we indeed see that the long-wavelength perturbation can be renormalized into the local spacetime coordinate consistently with the literature~\cite{Bravo:2017wyw,Bravo:2017gct,Bravo:2020hde}.
In fact, the solution $\xi^0=\zeta_\uL/H$ in the constant Hubble approximation kills the $(\ns-1)$ terms in our formula~\eqref{eq: modulated 2pt} and then the bispectrum evaluated at the time $t-\xi^0$ in order to fix the physical time $\bar{t}$ vanishes during the USR phase.
However it does not necessarily mean the disappearance of the physical squeezed bispectrum \emph{after} inflation.
If the non-attractor phase is followed by an attractor universe as a standard scenario, the local oberver effect on the squeezed bispectrum is again given by $-(1-\ns)P_\zeta(k_\uL)P_\zeta(k_\uS)$, but it cancels only the first term in our formula~\eqref{eq:gsfst} in fact.
While the last term represents the transition effect as discussed in the main part of the paper, it is interesting to note that the $(\ns-1)$ term also remains (though $\ns-1\to0$ in the exact USR limit), caused by the presence of $\xi^0$. It is understood as follows. During the USR phase, the physical time $\bar{t}$ corresponds with the uniform-$\dot{\phi}$ (and thus uniform-density) slice, but the uniform-density slice is given by the comoving time $t$ after the transition to the attractor phase. Then the time difference between them, $\xi^0$, yields extra expansion, i.e., modulation in physical scale. In fact an equal backward e-folds hypersurface from the end of inflation differently defines both $\phi$ and $\dot{\phi}$, depending on the phase-space trajectory, so that local observers differently see the local physics in each patch. That is why the $(\ns-1)$ term arises rather if the local observer effect is taken into account.

One may wonder why the transition from an USR phase to a SR phase is important for the later behavior of long-wavelength perturbations, while the transition from a SR phase to an oscillating phase after inflation is not. In order to answer this question, it should be noticed that, during a non-attractor (USR) phase, its dynamics is completely controlled solely by the momentum (velocity) of an inflaton but its phase-space trajectory does not converge to a single trajectory because it still depends on an initial field value. This is the key reason why the transition from an USR phase to a SR one can imprint the observed squeezed bispectrum, that is, local physics. On the other hand, during an attractor (SR) phase, its phase-space trajectory effectively converges to a single trajectory and hence the dynamics remains adiabatic in a later transition phase unless some chaotic behavior exaggerating (extremely) small difference among (converging) trajectories is present at the transition. Thus, such transition phase does not give important imprints.

\acknowledgments

We are grateful to 
Yi-Fu Cai,
Xingang Chen,
Hayato Motohashi,
Gonzalo A. Palma,
Spyros Sypsas,
and Dong-Gang Wang
for helpful discussions
T.~S. was supported by the MEXT Grant-in-Aid for Scientific Research on Innovative Areas No. 17H06359, and No. 19K03864. Y.~T. is supported by JSPS KAKENHI Grants No. JP18J01992 and No. JP19K14707. M.\,Y. is supported in part by JSPS Grant-in-Aid for Scientific Research Numbers 18K18764 and JSPS Bilateral Open Partnership Joint Research Projects.

\appendix

\section{\boldmath Uniform-$\dot{\phi}$ gauge}
\label{sec: uniform-dotphi}

In the main body, we start from the comoving gauge $\phi(t,\bfx)=\phi_0(t)$. The transformed field value is then automatically fixed to the background one as $\bar{\phi}(\bar{t})=\phi_0(t)$, so that the bispectrum correction is expressed by $P_\zeta$'s dependence on the remained DoF, $\dot{\phi}$.
However $\phi$ and $\dot{\phi}$ are originally equal, independent DoF and it should be possible to fix the momentum as $\bar{\phi}^\prime(\bar{t})=\dot{\phi}_0$ and express the bispectrum correction by the $\phi$-dependence of the power spectrum.
We clarify it in this appendix.

Let us start without specifying the spacetime gauge first:
\bae{
    \dd{s}^2=-\ee^{2\delta\calN_\uL}\dd{t}^2+a^2(t)\ee^{2\psi_\uL}(\dd{\bfx}+\bm{\calN}_\uL\dd{t})^2 \qc
    \phi(t,\bfx)=\phi_0(t)+\delta\phi_\uL(t).
}
In the long-wavelength limit, the linear constraint reads
\bae{
    \delta\calN_\uL(t)=-\epsilon_HH\delta u_\uL(t)+\frac{\dot{\psi}_\uL(t)}{H} \qc
    \bm{\calN}_\uL(t,\bfx)=\frac{1}{3}\epsilon_H\bfx\dot{\zeta}_\uL,
}
with the comoving curvature perturbation $\zeta_\uL=\psi_\uL+H\delta u_\uL$ and the velocity perturbation $\delta u_\uL=-\delta\phi_\uL/\dot{\phi}_0$.
We then look for a coordinate transformation
\bae{
    t\to\bar{t}=t+\xi^0(t) \qc
    \bfx\to\bar{\bfx}=\ee^{\beta(t)/3}\bfx \qc
    \phi(t)\to\bar{\phi}(\bar{t})=\phi(t)=\phi_0(t)+\delta\phi_\uL(t),
}
such that the transformed momentum coincides with the background one as
\bae{\label{eq: uniform dotphi}
    \bar{\phi}^\prime(\bar{t})=\pqty{1+\dot{\xi}_0}^{-1}\pqty{\dot{\phi}_0+\delta\dot{\phi}_\uL}=\dot{\phi}_0,
}
and $\bar{\phi}(\bar{t})$ satisfies the ``background-like" EoM
\bae{\label{eq: background-like EoM}
    \pdv[2]{\bar{\phi}}{\bar{t}}+3\bar{H}\pdv{\bar{\phi}}{\bar{t}}+\pdv{}{\bar{\phi}}V(\bar{\phi})=0 \qc
    3\Mpl^2\bar{H}^2=\frac{1}{2}\pqty{\pdv{\bar{\phi}}{\bar{t}}}^2+V(\bar{\phi}),
}
in the ``background-like" spacetime
\bae{\label{eq: vanishing metric ptb}
    \dd{s}^2=-\dd{\bar{t}}^2+\bar{a}^2(\bar{t})\dd{\bar{\bfx}}^2 \qc
    \bar{a}(\bar{t})=a(t)\ee^{\alpha(t)}.
}

First the vanishing metric perturbation condition~\eqref{eq: vanishing metric ptb} fixes the transformation parameters $\xi^0$, $\alpha$, and $\beta$ as
\bae{
    \dot{\xi}^0=\delta\calN_\uL \qc
    \alpha+\frac{1}{3}\beta=\psi_\uL \qc
    \dot{\beta}=\epsilon_H\dot{\zeta}_\uL.
}
The uniform-$\dot{\phi}$ condition~\eqref{eq: uniform dotphi} then reads
\bae{\label{eq: uniform dotphi gauge}
    \delta\dot{\phi}_\uL=\epsilon_HH\delta\phi_\uL+\frac{\dot{\phi}_0}{H}\dot{\psi}_\uL, \quad \Leftrightarrow \quad
    \psi_\uL=\zeta_\uL+\frac{1}{H\eta_1}\dot{\zeta}_\uL,
}
with a slow-roll parameter $\eta_1=\ddot{\phi}_0/(H\dot{\phi}_0)$.
As the right-hand side of the second equation is deterministic gauge-independently, it can be understood as a gauge condition onto the spatial curvature $\psi_\uL$.
Making use of EoM for $\zeta_\uL$ in the long-wavelength limit~\eqref{eq: zetaL EoM},
\bae{
    \dv{}{t}\pqty{\epsilon_Ha^3\dot{\zeta}_\uL}=0,
}
one can check that this gauge condition is in fact consistent with the ``background-like" EoM~\eqref{eq: background-like EoM}.
Therefore, if one starts from the time slice satisfying the gauge condition~\eqref{eq: uniform dotphi gauge} (instead of the comoving slice), one can indeed fix the momentum as $\bar{\phi}^\prime(\bar{t})=\dot{\phi}_0$ (instead of the field value itself).

In this gauge, the background trajectory is characterized by $\pqty{\bar{\phi}(\bar{t}),\bar{a}(\bar{t})}$. Thus the short-wavelength curvature perturbation is equivalent to the solution on this modulated trajectory, similarly to Eq.~\eqref{eq: zetaS by general solution}, as
\bae{
    \zeta_\uS(t,\bfx)=\zeta\bqty{\bar{t},\bar{\bfx}\mid\bar{\phi}(\bar{t}),\bar{a}(\bar{t})}.
}
At leading order in $\zeta_\uL$, it can be explicitly written as
\bae{
    \zeta_\uS(t,\bfx)\simeq\zeta\left[t+\xi^0,\pqty{1+\frac{\beta}{3}}\bfx\relmiddle{|}\phi_0(t+\xi^0)-\pqty{\xi^0-\frac{\dot{\zeta}_\uL}{H^2\eta_1}}\dot{\phi}_0,(1+\alpha-H\xi^0)a(t+\xi^0)\right].
}
Neglecting the time dependence of $(1+\alpha-H\xi^0)$ again, it reads
\bae{
    \zeta_\uS(t,\bfx)\simeq\zeta\left[t+\xi^0,\pqty{1+\zeta_\uL+\frac{\dot{\zeta}_\uL}{H\eta_1}-H\xi^0}\bfx\relmiddle{|}\phi_0(t+\xi^0)-\pqty{\xi^0-\frac{\dot{\zeta}_\uL}{H^2\eta_1}}\dot{\phi}_0,a(t+\xi^0)\right].
}
Hence the Fourier-space two-point function is given by
\bae{
    \braket{\zeta_\uS(\bfk_1)\zeta_\uS(\bfk_2)}&=\braket{\zeta(\bfk_1)\zeta(\bfk_2)}+\xi^0(\bfk_\uL)\dot{P}_\zeta[t,k_\uS\mid\phi_0] \nonumber \\
    &\quad-\pqty{\zeta_\uL(\bfk_\uL)+\frac{\dot{\zeta}_\uL(\bfk_\uL)}{H\eta_1}-H\xi^0(\bfk_\uL)}(\ns-1)P_\zeta[t,k_\uS\mid\phi_0] \nonumber \\
    &\quad-\pqty{\xi^0(\bfk_\uL)-\frac{\dot{\zeta}_\uL(\bfk_\uL)}{H^2\eta_1}}\dot{\phi}_0\partial_\phi P_\zeta[t,k_\uS\mid\phi]_{\phi=\phi_0}.
}
In terms of the bispectrum, one obtains
\bae{
    \lim_{k_3\to0}B_\zeta(t;k_1,k_2,k_3)&=(1-\ns)\pqty{P_\zeta(k_\uL)+\frac{1}{2H\eta_1}\dot{P}_\zeta(k_\uL)}P_\zeta(k_\uS) \nonumber \\
    &\quad-\pqty{\int^t\frac{3+\eta_1+\eta_2}{\eta_1}\frac{\partial_{t^\prime}P_\zeta(t,t^\prime;k_\uL)}{H}\dd{t^\prime}}\bqty{\dot{P}_\zeta(k_\uS)+H(\ns-1)P_\zeta(k_\uS)} \nonumber \\
    &\quad+\pqty{\frac{1}{2H^2\eta_1}\dot{P}_\zeta(k_\uL)+\int^t\frac{3+\eta_1+\eta_2}{\eta_1}\frac{\partial_{t^\prime}P_\zeta(t,t^\prime;k_\uL)}{H}\dd{t^\prime}}\dot{\phi}_0\partial_\phi P_\zeta[t,k_\uS\mid\phi]_{\phi=\phi_0},
}
with use of the equation for $\xi^0$ in this gauge,
\bae{
    \dot{\xi}^0=-\frac{3+\eta_1+\eta_2}{\eta_1}\frac{\dot{\zeta}_\uL}{H},
}
with $\eta_2=\dot{\eta}_1/(H\eta_1)$.
In the attractor limit, the conservation of $\zeta_\uL$ again leads to Maldacena's consistency relation~\eqref{eq: Maldacena CR}.
We also note that, in the exact USR limit, $\ns-1\to0,\eta_1\to-3,\eta_2\to0$, and also $\partial_\phi P_\zeta\to0$ due to the shift symmetry, this bispectrum completely disappears. This is because our time slice is not the comoving one and thus correlation functions of the time-evolving curvature perturbation can be different from those on the comoving slice.
If the USR phase is followed by a SR phase, even though curvature perturbations converge, $\dot{P}_\zeta\to0$, the term of $\phi$-dependence remains as
\bae{
    \lim_{k_3\to0}B_\zeta(k_1,k_2,k_3)\to\pqty{\int^t\frac{3+\eta_1+\eta_2}{\eta_1}\frac{\partial_{t^\prime}P_\zeta(t,t^\prime;k_\uL)}{H}\dd{t^\prime}}\dot{\phi_0}\partial_\phi P_\zeta[t,k_\uS\mid\phi]_{\phi=\phi_0},
}
and can give a non-zero contribution.

In the exact USR limit $\eta_1\to-3$, $\zeta$'s growth, $\dot{\zeta}=3H\zeta$, indicates that the gauge condition~\eqref{eq: uniform dotphi gauge} is nothing but the flat-slice rule $\psi_\uL=0$. As $\xi^0=0$ in this limit, $\bar{t}$ is also understood as a flat-slice time coordinate.
On the other hand, if the phase-space trajectory $(\bar{\phi},\bar{\phi}^\prime)$ converges to the background one $(\phi_0,\dot{\phi}_0)$ as an attractor solution, both $\bar{\phi}$ and $\bar{\phi}^\prime$ (and thus the local energy density $\bar{\rho}=\frac{1}{2}\bar{\phi}^\prime{}^2+V(\bar{\phi})$) simultaneously coincide with background values at the time $\bar{t}=t+\xi^0$.
Therefore, if inflation starts from an USR phase and then proceeds to a SR phase, $\xi^0$ is a time difference from an initial flat slice to a final uniform-density slice.
It reads $\zeta_\uL=H\xi^0$ according to the $\delta N$ formalism~\cite{Lyth:2004gb}.
Making use of this and noting that the initial condition for $\bar{\phi}$ is taken as
\bae{
    \bar{\phi}(t_\uL)=\phi_0(t_\uL)+\delta\phi_\uL \qc
    \bar{\phi}^\prime(t_\uL)=\dot{\phi}_0,
}
also in this gauge similarly to the comoving case~\eqref{eq: def of delta phi}, one successfully reproduces our formula~\eqref{eq: Suyama formula},
\bae{
    \fNL\to\frac{5n}{4Z_\uc} \qc
    Z_\uc=1+3H\dot{\phi}_\uc^2\int^\infty_{t_\uc}\frac{\exp\pqty{\int^t_{t_\uc}F_{\dot{\phi}}\dd{t^\prime}}}{\dot{\phi}_0^2(t)}\dd{t},
}
following the same procedure.

\bibliography{main}
\end{document}